\documentclass[preprint]{iucr}              

     \journalcode{J}              
                                  
\usepackage{siunitx} 
\usepackage[table]{xcolor}
\definecolor{tableShade}{gray}{0.9}
\usepackage{tabularx}
\usepackage{mathtools}
\usepackage{amssymb}
\usepackage{amsmath}
\usepackage[version=4]{mhchem}
\usepackage{siunitx}
\usepackage{gensymb}

\begin{document}




\title{Investigating model influence on the analytical resolution of neutron reflectometry}


\cauthor[a,b]{Nicolas}{Shiaelis}{nicolas.shiaelis@st-annes.ox.ac.uk}{address if different from \aff}
\author[b]{Luke A.}{Clifton}
\cauthor[c,d]{Andrew R.}{McCluskey}{andrew.mccluskey@bristol.ac.uk}

\aff[a]{Biological Physics Research Group, Clarendon Laboratory, Department of Physics, University of Oxford, Oxford, OX1 3PU \country{United Kingdom}}
\aff[b]{ISIS Neutron and Muon Source, Rutherford Appleton Laboratory, Didcot, Oxfordshire OX11 0QX, \country{United Kingdom}}
\aff[c]{School of Chemistry, University of Bristol, Cantock's Close, Bristol, BS8 1TS, \country{United Kingdom}}
\aff[d]{European Spallation Source ERIC, Ole Maaløes vej 3, 2200 København N, \country{Denmark}}









\maketitle                        


\begin{abstract}
Neutron reflectometry is a critical tool for investigating the structure of thin films and interfaces. 
However, the misapplication of the Born approximation to reflection geometry leads some to assume that the minimum thickness that may be probed by neutron reflectometry is limited by the $Q$-range of the measurement. 
In this study, we use model-dependent analysis, multiple isotopic contrasts, and magnetic spin states, to show that it is possible to resolve structures significantly small than this perceived limit. 
To quantify this ``analytical resolution'', we employ Bayesian model selection, offering a robust and quantifiable comparison between different analytical models. 
We believe that this work offers pivotal insights for the analysis of neutron reflectometry and hope that it will contribute to more accurate and information-rich analyses in the future.
\end{abstract}

\section{Introduction}

In the elastic scattering regime, it is commonly accepted that the minimum resolvable length scale, $L_{\mathrm{min}}$, is defined by the maximum measured scattering vector, $Q_{\mathrm{max}}$, 
\begin{equation}
    L_{\mathrm{min}} = \frac{2\pi}{Q_{\mathrm{max}}}.
    \label{eqn:limit}
\end{equation}
This limit arises from the observation that the differential cross section for elastic scattering is related to the scattering length density (SLD) profile through a Fourier transform, the magnitude of which will decay to zero at scattering vectors of $2\pi / L_{\mathrm{min}}$. 
The observation, itself, is dependent on the Born approximation \cite{born1926quantenmechanik}, which assumes that there is a single scattering event between the probing radiation and atoms which scatter them. 

Neutron and X-ray reflectometry are elastic scattering techniques that are important in the study of interfacial structures, such as model membranes~\cite{Clifton_envelope,Hughes}, surfactant monolayers~\cite{McCluskey2019}, and magnetic multilayers~\cite{Chen2022}.
As with other reciprocal space measurements, the analysis of reflectivity data presents a significant challenge due to the phase problem~\cite{phase-problem}. 
Frequently, it is assumed that reflectivity measurements are limited in resolvable resolution by the same Eqn.~\ref{eqn:limit}.
It is well documented~\cite{sivia_2011_elementary,Sears1993}, however, that the Born approximation does not hold for reflectometry measurements. 
The reason that the Born approximation does not hold for reflectometry appears to be a combination of multiple factors, typically related to the scattering geometry of the measurement. 

Measurements governed by the Born approximation can be analysed by an inverse Fourier transform, e.g., Fourier inversion of small angle scattering~\cite{Glatter1977}.
The same approach has been applied to reflectometry data~\cite{Majkrzak1998,Li1996,Bridou1994}, overlooking the collapse of the Born approximation already discussed. 
The more common approach to reflectivity data analysis is a model-dependent approach~\cite{McCluskey2020}, where a model system is proposed and model reflectivity calculated, using the optical matrix formalism, and the model is then refined against the experimental data.

It is still regularly suggested that the maximum $Q$-value measured limits the length scale probed by reflectometry, despite evidence that the Born approximation does not apply in the reflection geometry.
In this work, we show, analytically, that this length scale limitation does not hold in the analysis of neutron reflectometry measurements when a model-dependent approach is used.
To achieve this, we establish an analysis resolution metric, through the robust analysis of neutron reflectivity data by nested sampling~\cite{skilling}.
This metric quantifies the minimum length scale, for a given model parameter, that can be observed. 

We concentrate on a floating bilayer system~\cite{Clifton_Self_assembled}, a widely used and well-understood model system for studying lipid membranes.
It is shown that the analytical resolution, instead of depending on the maximum $Q$-value, is related to both the data and the analytical model. 
In particular, we investigated the effect of isotopic/spin contrasts and model assumptions on the analytical resolution. 
By providing a better understanding of the influence of the collected data and the analytical model on the resolution, we believe this work can inform the development of more comprehensive models for interpreting neutron reflectivity data. 
Specifically, our analytical methodology, which demonstrates the enhanced resolution achievable through model-dependent analysis, can guide the refinement of these models. 
Furthermore, our findings can aid in the design of experiments by offering insights into how data collection and model assumptions impact the minimum resolvable length scale.
\section{Materials and Methods}

\subsection{Materials}

DPPC (1,2-dipalmitoyl-\emph{sn}-glycero-3-phosphocholine) was obtained from Avanti polar lipids (Alabaster, AL, USA) and used without further purification. 
The self-assembled monolayer material (TAAA-SAM; \ce{HS-CH2-(CONH)15-CH2(OH)2}) was obtained from Prochimia surfaces (Gdansk, Poland). 
Deuterium oxide (\ce{D2O}), HEPES buffer salts, and all other
chemicals were sourced from Sigma-Aldrich or Fisher Scientific (Loughborough, UK). 
Silicon substrates were obtained from Crystran (Poole, UK). 

\subsection{Sample preparation}
\begin{figure}
    \centering
    \includegraphics{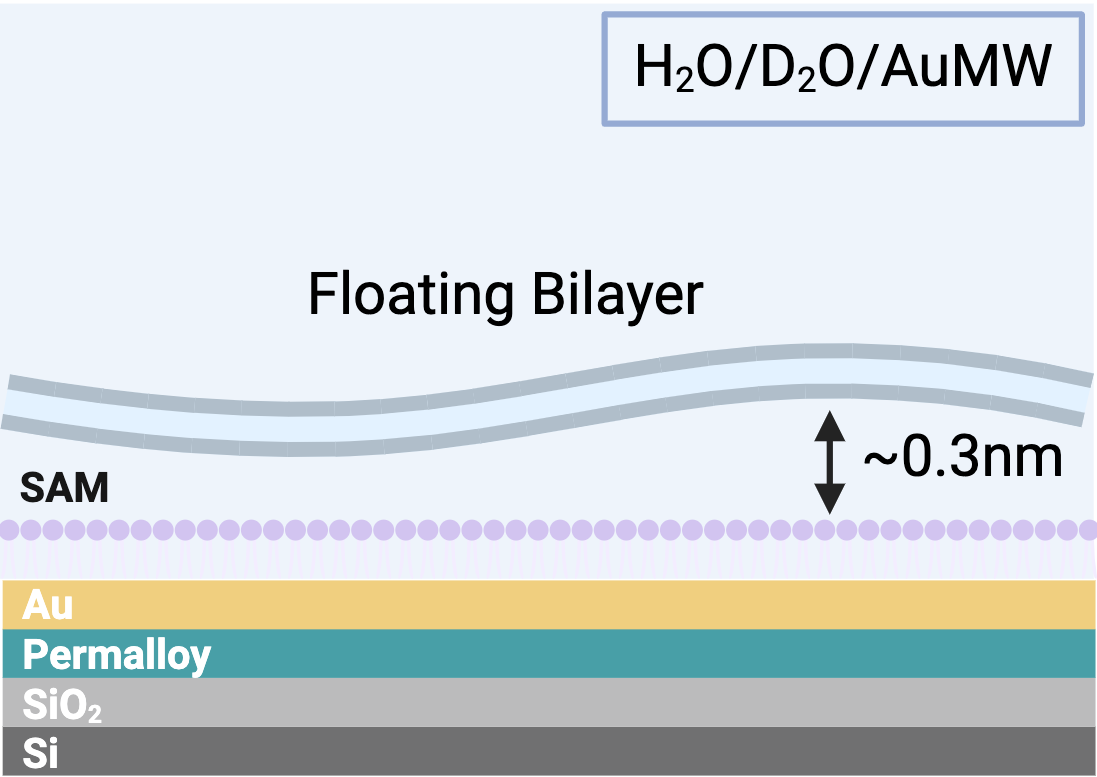}
    \caption{Schematic of the sample that was measured, showing the construction of thefloating bilayer system.}
    \label{fig:system schematic}
\end{figure}
In this work, we focus on a floating bilayer system, which has been extensively studied~\cite{Clifton_Self_assembled, fragneto2001fluid,daillant2005structure}. 
The floating bilayer system was prepared as in the work of Clifton \emph{et al.}~\cite{Angewandte_Clifton}.
The resulting sample consisted of a substrate of silicon/silicon oxide with a permalloy and a gold layer.
On top of this substrate, the self-assembled moonolayer of the TAAA-SAM material was deposited. 
The floating bilayer was then formed adjacent to this by Langmuir-Blodgett/Langmuir-Schaefer deposition (Fig.~\ref{fig:system schematic}).
The system was measured under solution in custom-built solid/liquid flow cells.

\subsection{Neutron reflectometry measurements}
\begin{table}[]
\begin{tabular}{l | cc | cc | ccc}
\hline
\textbf{} &
  \multicolumn{2}{c}{\textbf{Sample}} &
  \multicolumn{2}{| c |}{\textbf{\begin{tabular}[c]{@{}c@{}}Permalloy \\ spin state\end{tabular}}} &
  \multicolumn{3}{c}{\textbf{Water}} \\ \hline
\textbf{Contrast} &
  \textbf{Substrate} &
  \textbf{\begin{tabular}[c]{@{}c@{}}Substrate \\ + bilayer\end{tabular}} &
  \textbf{po} &
  \textbf{mo} &
  \textbf{H2O} &
  \textbf{D2O} &
  \textbf{AuMW} \\ \hline
\rowcolor[HTML]{EFEFEF} 
\ce{H2O}/po         & • &   & • &   & • &   &   \\
\ce{H2O}/mo         & • &   &   & • & • &   &   \\
\rowcolor[HTML]{EFEFEF} 
\ce{D2O}/po         & • &   & • &   &   & • &   \\
\ce{D2O}/mo         & • &   &   & • &   & • &   \\
\rowcolor[HTML]{EFEFEF} 
\ce{H2O}/po/h-DPPC  &   & • & • &   & • &   &   \\
\ce{H2O}/mo/h-DPPC  &   & • &   & • & • &   &   \\
\rowcolor[HTML]{EFEFEF} 
\ce{D2O}/po/h-DPPC  &   & • & • &   &   & • &   \\
\ce{D2O}/mo/h-DPPC  &   & • &   & • &   & • &   \\
\rowcolor[HTML]{EFEFEF} 
AuMW/po/h-DPPC &   & • & • &   &   &   & • \\
AuMW/mo/h-DPPC &   & • &   & • &   &   & • \\ \hline
\end{tabular}
\caption{The isotopic/spin contrasts that gave rise to the ten neutron reflectometry measurements that make up the data set investigated herein.}
\label{table:contrasts explained}
\end{table}
Polarised neutron reflectometry (PNR) measurements where conducted at the POLREF reflectometer at the ISIS Neutron and Muon Source. 
This instrument measures the reflection of a white neutron beam and examines the reflection of a single neutron spin state from the sample.
Polarisation was achieved using a polarizing mirror and a spin flipper. 
Reflectivity data was gathered across a $Q_z$ range of approximately \SIrange{0.01}{0.3}{\per\angstrom} (at \SI{2.5}{\percent} resolution) using glancing angles of \SIlist{0.25;0.5;1.25;2.5}{\degree} with neutron wavelengths of \SIrange{2}{12}{\angstrom}. 
The total illuminated sample length was $~60 mm$.

The floating bilayer system was analyzed using six distinct isotopic/spin contrasts, and an additional four contrasts were measured for the substrate. This was accomplished by adjusting the magnetic orientation of the permalloy layer with an external magnetic field, which allowed the neutron spin to align either parallel (p) or anti-parallel (m) to the field. Furthermore, the surrounding medium was varied using either \ce{H2O},\ce{D2O}, or gold-contrast matched water (AuMW) to achieve different contrast conditions (Tab.~\ref{table:contrasts explained}). A liquid chromatography pump (Knauer Smartline 1000) was connected to the liquid cell inlet for programmable control of the H2O/D2O solution mixture in the sample cell.

\subsection{Analytical model}

The model-dependent analysis of neutron reflectivity data is performed by applying the Abel\`{e}s optical matrix formalism~\cite{Abels1950} to a structure of layers with given scattering length densities and thicknesses with interfacial roughnesses between the layers~\cite{Nvot1980}.
Here, we parameterise an analytical model that describes the floating bilayer system that consists of the layers shown in Figure~\ref{fig:system schematic}.
Table~\ref{table:parameters+distributions} gives the parameter prior distributions for those that were allowed to vary. 
\begin{table}[]
\begin{tabular}{llcccc}
\hline
   & Parameter                   & Prior Range           & Prior distribution & Mean & Std \\ \hline
\rowcolor[HTML]{EFEFEF} 
1  & SiO2 SLD                    & {[}3.41, 7.11{]}      & Truncated normal   & 4    & 2   \\
2  & SiO2 thickness              & {[}12.00, 18.00{]}    & Truncated normal           & 15   & 1   \\
\rowcolor[HTML]{EFEFEF} 
3  & SiO2 roughness              & {[}12.00, 18.00{]}    & Truncated normal           & 15   & 1   \\
4  & Permalloy spin up SLD       & {[}9.00, 12.00{]}     & Uniform            & -    & -   \\
\rowcolor[HTML]{EFEFEF} 
5  & Permalloy spin down SLD     & {[}6.00, 10.00{]}     & Uniform            & -    & -   \\
6  & Permalloy thickness         & {[}100.00, 200.00{]}  & Uniform            & -    & -   \\
\rowcolor[HTML]{EFEFEF} 
7  & Permalloy roughness         & {[}5.00, 11.00{]}     & Uniform            & -    & -   \\
8  & Au thickness                & {[}100.00, 200.00{]}  & Uniform            & -    & -   \\
\rowcolor[HTML]{EFEFEF} 
9  & Au roughness                & {[}4.00, 8.00{]}      & Uniform            & -    & -   \\
10 & SAM Area Per Molecule (APM) & {[}15.00, 30.00{]}    & Truncated normal           & 23   & 1.5 \\
\rowcolor[HTML]{EFEFEF} 
11 & SAM tail thickness          & {[}15.00, 20.00{]}    & Uniform            & -    & -   \\
12 & SAM head thickness          & {[}7.00, 11.00{]}     & Truncated normal           & 9    & 0.7 \\
\rowcolor[HTML]{EFEFEF} 
13 & SAM hydrtation              & {[}0.00, 1.00{]}      & Truncated normal           & 0.5  & 1.5 \\
14 & Water Interlayer thickness  & {[}0.10, 5.00{]}      & Truncated normal         & 0.5  & 1   \\
\rowcolor[HTML]{EFEFEF} 
15 & Bilayer tail volume         & {[}600.00, 1000.00{]} & Uniform            & -    & -   \\
16 & Bilayer head volume         & {[}300.00, 380.00{]}  & Uniform            & -    & -   \\
\rowcolor[HTML]{EFEFEF} 
17 & Bilayer defect hydration    & {[}0.00, 1.00{]}      & Truncated normal   & 0.1  & 0.5 \\
18 & Bilayer head hydration      & {[}0.10, 1.00{]}      & Truncated normal   & 0.5  & 1   \\ \hline
\end{tabular}
\label{table:parameters+distributions}
\caption{A list of all the parameters used in the model with the corresponding prior range, and prior distribution.}
\end{table}

The substrate layers shown in Figure~\ref{fig:system schematic} were described with thicknesses (with the exception of the semi-infinite silicon layer) and scattering length densities and interfacial roughnesses between each. 
The SAM layer is split into separate heads and tails layers, and the scattering length densities, $\beta$, for these layers were constrained based on the area per molecule (APM) of the SAM with the following relation
\begin{equation}
    \beta = \frac{b}{d \mathrm{APM}} + \beta_{\mathrm{sol}} (1 - \phi),
\end{equation}
where, $b$ is the sum of the scattering lengths of the atoms that make up either the head or the tail, $d$ is the thickness of the head or tail layer, $\phi$ is the volume fraction of the head or tail material in the solvent of scattering length density $\beta_{\mathrm{sol}}$. \
By having the same APM for both the head and tail, the chemical constraint that there are the same numbers of head and tail groups is achieved. 
The DPPC bilayer is described using two pairs of similar components, placed such that the tail layers are adjacent. 
However, unlike the SAM layer, the components are described in terms of the molecular volumes for the head and tail groups. 
This means that in order to introduce the chemical constraint to ensure an equal number of head and tail groups the tail thickness, $d_{\mathrm{tail}}$ is constrained as follows,
\begin{equation}
    d_{\mathrm{tail}} = \frac{d_{\mathrm{head}}\phi_{\mathrm{h}}V_{\mathrm{tail}}}{\phi_{\mathrm{tail}}V_{\mathrm{head}}},    
\end{equation}
where, $d{\mathrm{head}}$, $\phi_{\mathrm{head}}$, and $V_{\mathrm{head}}$ are the thickness, volume fraction, and molecular volume of the head group respectively, and similarly for the tail group parameters. 
Between the SAM and bilayer, there is a narrow interlayer of water and above the bilayer a semi-infinite water layer.

In addition, interfacial roughness, $\sigma$, is present at each of the three interfaces (solvent-head, head-tail, and tail-air), and is modeled using an error function. Notably, in the study by Campbell et al, the roughness was assumed to be conformal, such that it does not vary between interfaces. This assumption is reasonable in the case of a monolayer of a single lipid type which is the case for our system.

We have also used the area per molecule assumption, a common simplifying assumption used in the study of monolayers and thin films. It assumes that all molecules in the monolayer or thin film have the same cross-sectional area and that the total area of the monolayer or thin film is simply the product of the number of molecules and their individual cross-sectional area.

The area per molecule assumption is a useful simplification because it allows for straightforward calculations of the thickness and packing density of the monolayer or thin film from measurements of its total area and the number of molecules present.

\subsection{Bayesian model selection framework}




Bayesian model selection is a statistical framework used to compare and select the most suitable model from multiple options based on how well they fit the observed data. This process involves evaluating the relative likelihoods of different models, considering their complexity, and then choosing the model with the highest posterior probability.

This selection process is based on Bayes' theorem, which provides a way to calculate the posterior distribution - the probability of the model parameters given the observed data. The theorem combines the likelihood of the data given the model parameters and the prior probability of the parameters.

The evidence for the data given in our model, which is integral over all possible parameter combinations, is a crucial part of this calculation. This evidence can be efficiently estimated using nested sampling\cite{skilling}, a Monte Carlo method. Nested sampling\cite{ashton2022nested} provides a measure of how well the model fits the data, taking into account its complexity.

We then use Bayes factors\cite{jeffreys1998theory}, a popular method for comparing the relative likelihoods of different models in the Bayesian model selection framework. The Bayes factor between the two models is the ratio of their evidence, calculated on the same dataset. It is important to note that in the case where our \textit{a priori} belief in the two models is equal, the evidence ratio, called the Bayes factor, completely specifies the relative probability of the two models.

Considering a case where we have two competing models $M_1$ and $M_2$, both of which describe our system. Given some data $\textbf{D}$ the Bayes factor is defined as \cite{sivia2006data},
\begin{equation}
Bayes Factor (BF) = \frac{P(M_1|D,I)}{P(M_2|D,I)} = \frac{\mathcal{Z}_1}{\mathcal{Z}_2}\frac{P(M_1|I)}{P(M_2|I)}.
\end{equation}

\begin{figure}
    \centering
    \includegraphics[width=1\textwidth]{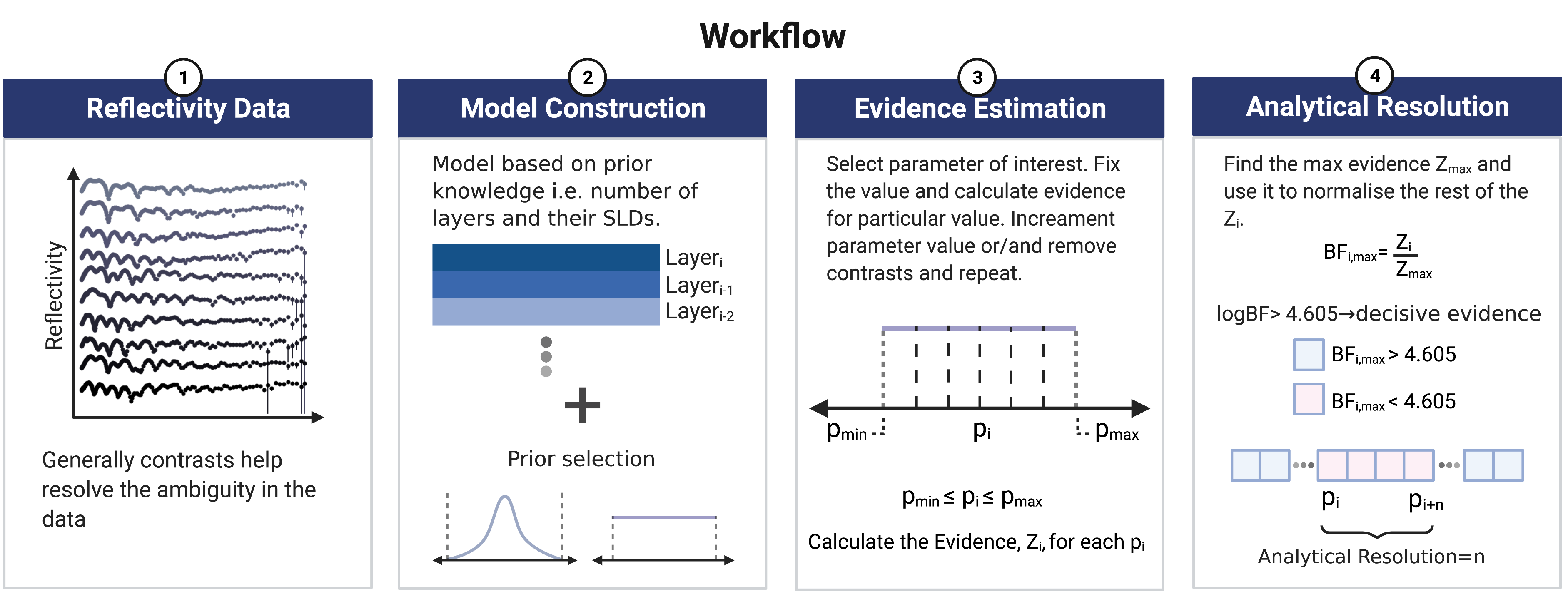}
    \caption{\textbf{Analytical Resolution Estimation Workflow.}\\ 
    {\large\textcircled{\small{1}}} Optimally good data set of a well-understood system, with $H_{2}O$, $D_{2}O$, $AuMW$ and spin contrasts.\\
    {\large\textcircled{\small{2}}} Code model creating layers with SLD, thickness and roughness value-ranges found in the literature, using all or part of the contrasts available.\\
    {\large\textcircled{\small{3}}} Estimate the evidence for incremental values of the parameter of interest.\\
    {\large\textcircled{\small{4}}} Analytical resolution calculation based on bayes factors.\\
    }
    \label{fig:Workflow}
\end{figure}

We select a parameter of interest, in this case the bilayer tail thickness, and run the nested sampling algorithm for a range of possible values, resulting in the evidence values associated with every parameter value. It should be noted that two models can only be compared for the same data (Fig. \ref{fig:Workflow} {\large\textcircled{\small{3}}}).
In this work we are working with log evidence values and every evidence value is normalised by the highest evidence for the particular set of data. This results in the logarithmic Bayes factors Fig. \ref{fig:Workflow} {\large\textcircled{\small{3}}}). 

\begin{equation}
\ln{(BF_{i,max})} = \ln{(\frac{\mathcal{Z}_i}{\mathcal{Z}_{max}})}=\ln{(\mathcal{Z}_i)}-\ln{(\mathcal{Z}_{max})}
\end{equation}

This allows us to define the analytical resolution of the technique as the number of consecutive parameter values with a Bayes factor smaller than $4.61$ (which is the most conservative value we can choose, see Table \ref{table:Bayes}). This definition allows us to quantify the analytical resolution and it is a metric of the effect of the chosen model (external information about the system) on the resulting resolution that accounts for both the data and the model.

\begin{table}
\begin{center}
\caption{Interpretation of Bayes factor, $BF_{i,max}$ between the $i^{th}$ and highest evidence model.}

\begin{tabular}{ccc}      

 \hline $BF_{i,max}$    & $\ln(BF_{i,max})$ & Interpretation \\ \hline
 $>100$       & $>4.61$             & Decisive evidence for $p_{max}$ \\
 $30-100$     & $3.40-4.61$       & Very strong evidence for $p_{max}$ \\
 $10-30$      & $2.30-3.40$      & Strong evidence for $p_{max}$ \\
 $3-10$       & $1.10-2.30$      & Anecdotal evidence for $p_{max}$ \\
 $1-3$        & $0-1.10$         & Not worth more than a mention \\ \hline
\end{tabular}
\label{table:Bayes}
\end{center}
\end{table}

\section{Results}

\subsection{Neutron reflectometry data analysis}
\begin{figure}
    \centering
    \includegraphics[width=1\textwidth]{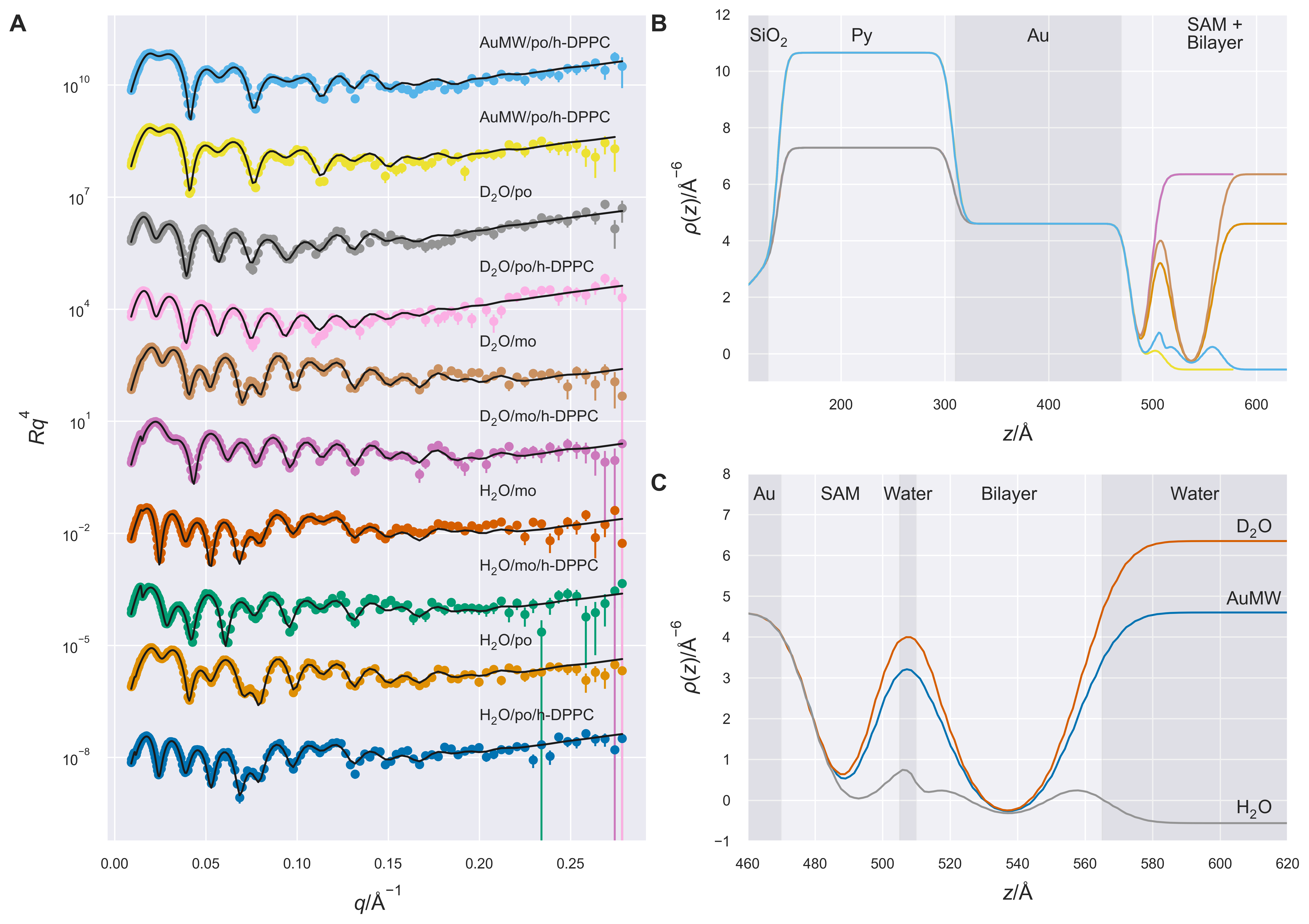}
    \caption{Reflectivity curve and corresponding SLD profiles.\textbf{(A)} The experimental (coloured lines) reflectometry and the median values for the model with the greatest evidence (black lines) for the tail thickness parameter. The different contrasts are offset by an order of magnitude in reflected intensity. \textbf{(B)} SLD profile for the substrate and the floating bilayer system. \textbf{(C)} Zoomed in SLD profile for the SAM and floating bilayer.}
    \label{fig:ReflectivityCurve}
\end{figure}

Figure \ref{fig:ReflectivityCurve} (A) presents a comparison between the experimental data and the reflectometry profiles obtained for the tail thickness parameter value with the highest evidence when all contrasts are utilised in the analysis. Across all contrasts, a clear agreement is evident between the experimental data and the model predictions. This observation confirms the validity of the assumptions discussed in Section 2.4.

\subsection{Effective resolution for a Single Dataset}

\begin{figure}
    \centering
    \includegraphics[width=1\textwidth]{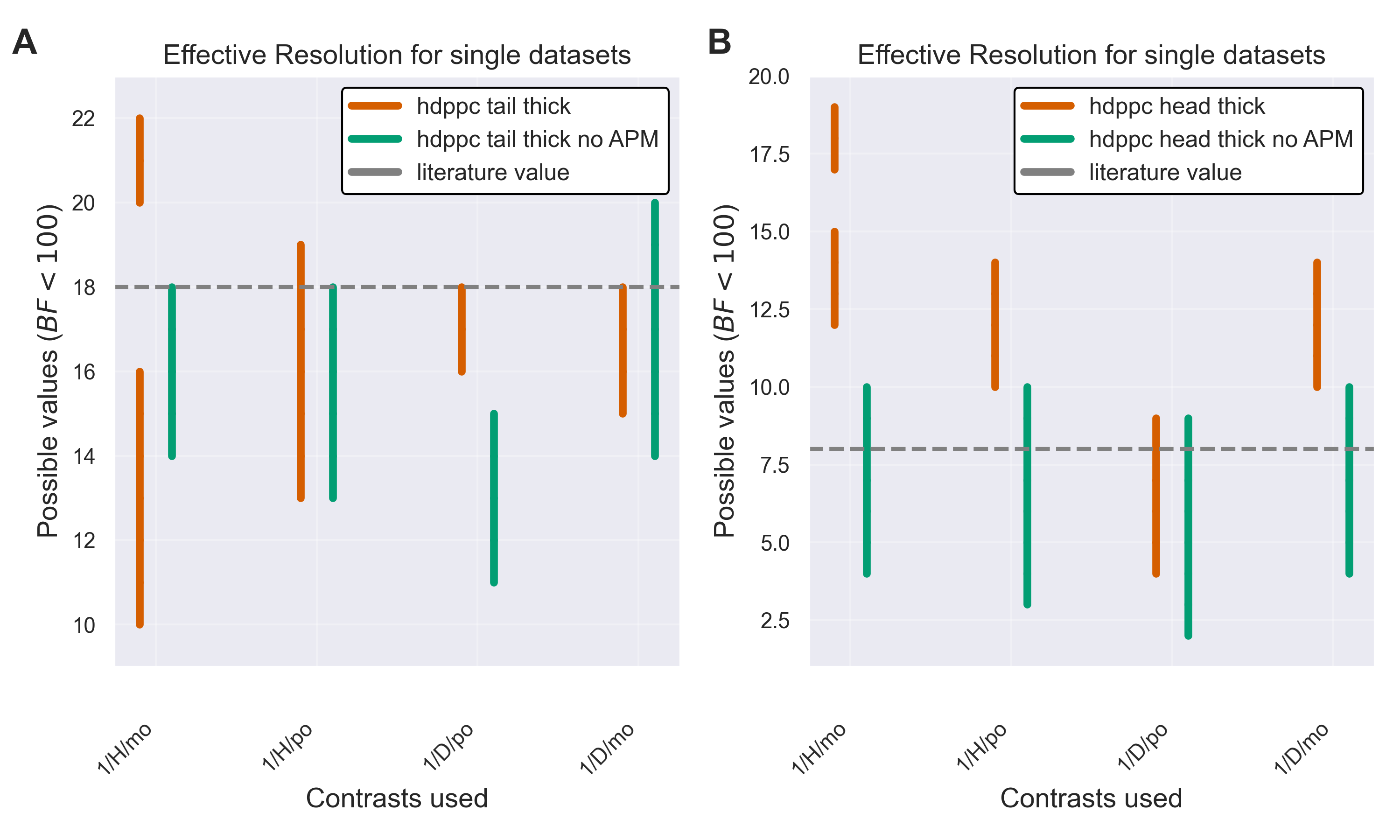}
    \vspace{-3em}
    \caption{Plots of the Bayes factors below the decisive evidence value for single contrasts. \textbf{(A)} Indecisive evidence plot for the bilayer tail thickness and \textbf{(B)} the bilayer head thickness with and without the APM constraint.}
    \label{fig:single-contrasts}
\end{figure}

Our investigation into the effective resolution of neutron reflectometry for a floating bilayer system illustrates the potential for empirical data to surpass the boundaries dictated by theoretical approximations. As can be seen in figure \ref{fig:single-contrasts} for the tail parameter for a single contrast the worst effective resolution achieved is \SI{10}{\angstrom} and for the head parameter \SI{7}{\angstrom} which in both cases is better than the theoretical limit. For reference the literature value for the tail thickness is \SI{18}{\angstrom} and for the head thickness is \SI{8}{\angstrom}\cite{vacklin2005composition}.

 The complex nature of the system under study, the occurrence of multiple scattering events, and the application of model fitting to the experimental data all contribute to an effective resolution that exceeds the limit suggested by the Born approximation (see equation \ref{eqn:limit}). 
 
\subsection{Effect of increasing contrasts on the effective resolution}

\begin{figure}
    \centering
    \includegraphics[width=1\textwidth]{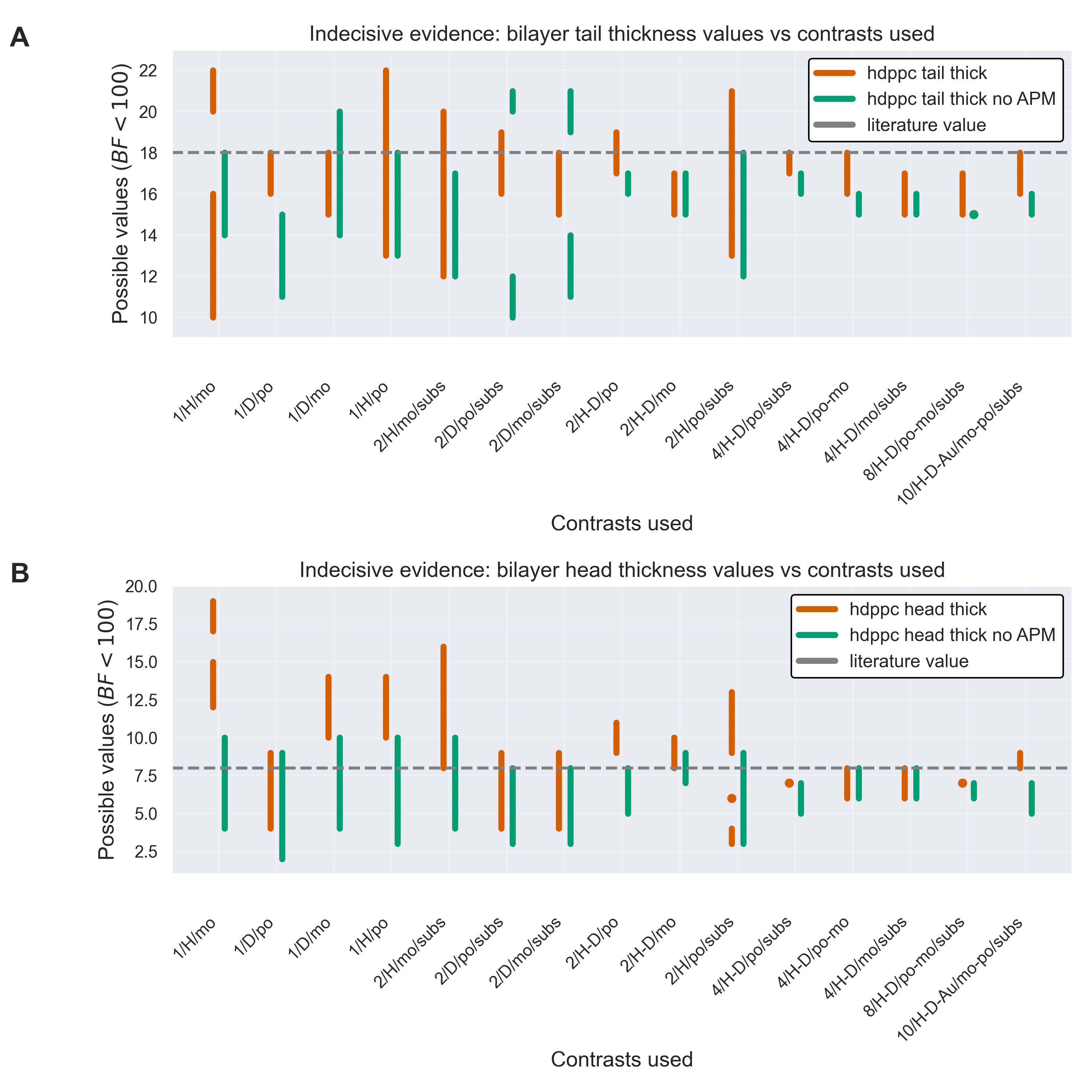}
    \vspace{-3em}
    \caption{Plots of the Bayes factors below the decisive evidence value for each combination of contrasts. \textbf{(A)} Indecisive evidence plot for the bilayer tail thickness and \textbf{(B)} the bilayer head thickness with and without the APM constraint.}
    \label{fig:APM-and-without}
\end{figure}
The analytical resolution for both the tail and head thickness parameters (see Figure \ref{fig:APM-and-without}) improves for an increasing number of contrasts. This makes sense intuitively as with more isotropic contrasts the parameter space gets smaller. This is true for both the APM and not APM restricted model.

\begin{table}[]
\resizebox{\textwidth}{!}{%
\begin{tabular}{lcccccccc}
\hline
\textbf{} &
  \multicolumn{1}{l}{\textbf{}} &
  \multicolumn{2}{c}{\textbf{Sample}} &
  \multicolumn{2}{l}{\textbf{\begin{tabular}[c]{@{}l@{}}Permalloy \\ spin state\end{tabular}}} &
  \multicolumn{3}{c}{\textbf{Water}} \\ \hline
\textbf{Name} &
  \textbf{\begin{tabular}[c]{@{}c@{}}Num. \\ contrasts\end{tabular}} &
  \textbf{Substrate} &
  \textbf{\begin{tabular}[c]{@{}c@{}}Substrate \\ + bilayer\end{tabular}} &
  \textbf{Up} &
  \textbf{Down} &
  \textbf{H2O} &
  \textbf{D2O} &
  \textbf{AuMW} \\ \hline
\rowcolor[HTML]{EFEFEF} 
\begin{tabular}[c]{@{}l@{}}10/H-D-Au/po-mo/subs\end{tabular}               & 10 & • & • & • & • & • & • & • \\
8/H-D/po-mo/subs                                                                & 8  & • & • & • & • & • & • &   \\
\rowcolor[HTML]{EFEFEF} 
\begin{tabular}[c]{@{}l@{}}4/H-D/po/subs\end{tabular}            & 4  & • & • & • &   & • & • &   \\
\begin{tabular}[c]{@{}l@{}}4/H-D/mo/subs\end{tabular}          & 4  & • & • &   & • & • & • &   \\
\rowcolor[HTML]{EFEFEF} 
\begin{tabular}[c]{@{}l@{}}4/H-D/po-mo\end{tabular}             & 4  &   & • & • & • & • & • &   \\
\begin{tabular}[c]{@{}l@{}}2/H/po/subs\end{tabular}                 & 2  & • & • & • &   & • &   &   \\
\rowcolor[HTML]{EFEFEF} 
\begin{tabular}[c]{@{}l@{}}2/D/po/subs\end{tabular}                 & 2  & • & • & • &   &   & • &   \\
\begin{tabular}[c]{@{}l@{}}2/H-D/po \end{tabular}   & 2  &   & • & • &   & • & • &   \\
\rowcolor[HTML]{EFEFEF} 
\begin{tabular}[c]{@{}l@{}}2/H/mo/subs\end{tabular}               & 2  & • & • &   & • & • &   &   \\
\begin{tabular}[c]{@{}l@{}}2/D/mo/subs\end{tabular}               & 2  & • & • &   & • &   & • &   \\
\rowcolor[HTML]{EFEFEF} 
\begin{tabular}[c]{@{}l@{}}2/H-D/mo\end{tabular} & 2  &   & • &   & • & • & • &   \\
\begin{tabular}[c]{@{}l@{}}1/H/po\end{tabular}         & 1  &   & • & • &   & • &   &   \\
\rowcolor[HTML]{EFEFEF} 
\begin{tabular}[c]{@{}l@{}}1/D/po\end{tabular}         & 1  &   & • & • &   &   & • &   \\
\begin{tabular}[c]{@{}l@{}}1/H/mo\end{tabular}       & 1  &   & • &   & • & • &   &   \\
\rowcolor[HTML]{EFEFEF} 
\begin{tabular}[c]{@{}l@{}}1/D/mo\end{tabular}       & 1  &   & • &   & • &   & • &   \\ \hline
\end{tabular}}
\caption{Details of the combinations of contrasts used in the analysis.}
\label{table:contrasts-info}
\end{table}

\subsection{Effect of removing APM assumption from the model}

For the tail thickness parameter (Figure \ref{fig:APM-and-without} A) the average value for most contrast combinations is higher when the APM assumption is coded into the model. On the other hand, the analytical resolution is worse when the APM assumption is used. This could be because the head thickness is calculated from the APM and tail thickness parameter when the APM assumption is used. Since the tail thickness is a well-encoded parameter whereas the head thickness is a poorly encoded parameter calculating one of the two combines their uncertainties. This is supported by Figure \ref{fig:APM-and-without} B, where we can see that the analytical resolution is better when the APM assumption is used. For both parameters, the average values estimated by the model are closer to the values reported in the literature\cite{vacklin2005composition} when the APM assumption is used.

\subsection{Comparison of H2O vs D2O contrasts}

\begin{figure}
    \centering
    \includegraphics[width=1\textwidth]{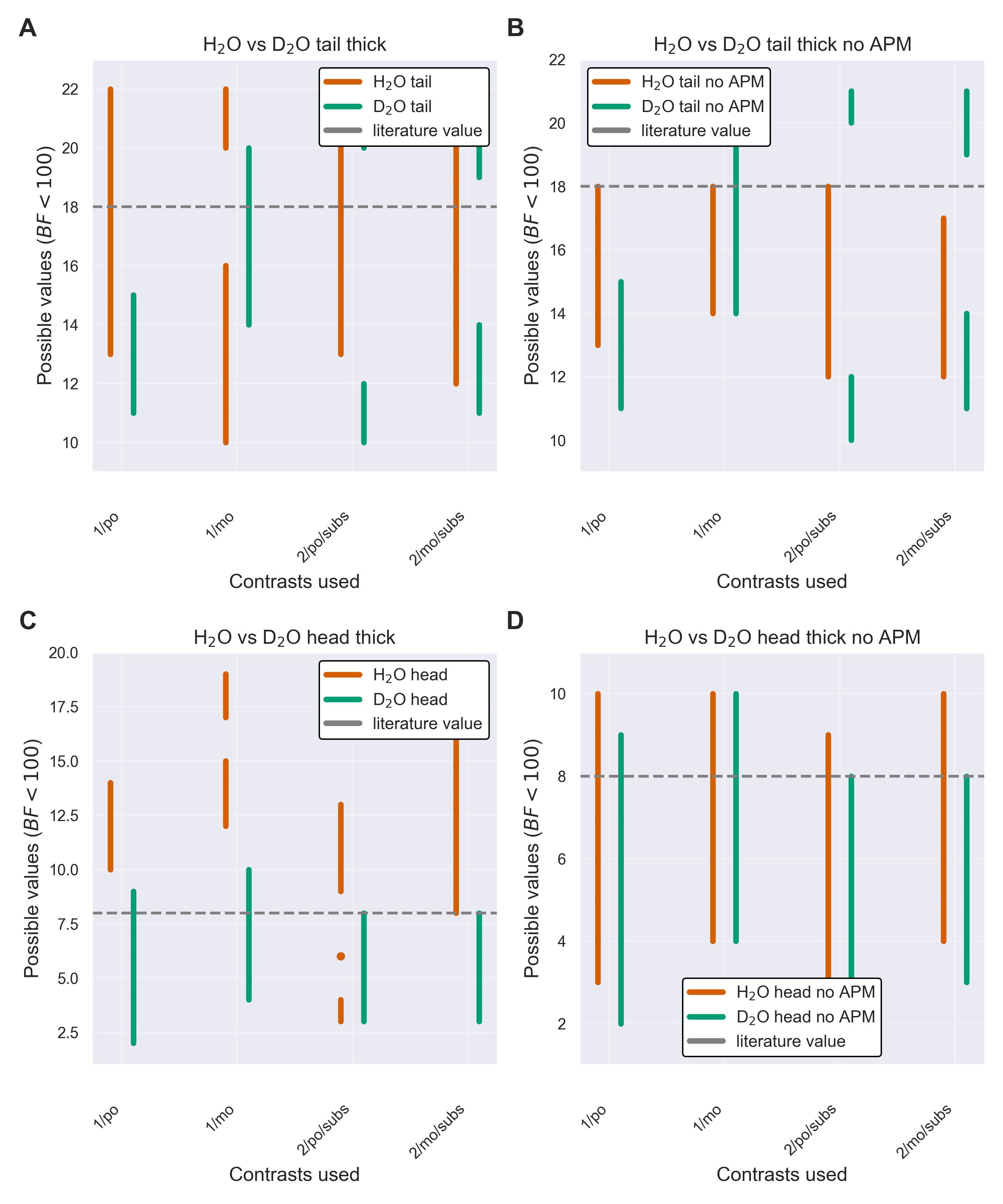}
    \vspace{-3em}
    \caption{Comparison plots of the Bayes factors below the decisive evidence value for $H_2O$ vs $D_2O$ contrasts only. Indecisive evidence plot for the \textbf{(A)} bilayer tail thickness with the APM constraint and \textbf{(B)} without, \textbf{(C)} the bilayer head thickness with the APM constraint and \textbf{(D)} without.}
    \label{fig:water-comparison}
\end{figure}

The analytical resolution for both the tail and head thickness parameters (see Figure \ref{fig:water-comparison} A and C) is generally better for $D_2O$ than $H_2O$ whereas when the APM assumption is not used the resolution is broadly similar for both $D_2O$ and $H_2O$ contrasts. The possible values for $D_2O$ and $H_2O$ are different likely due to the difference in hydration parameters for the two cases.

\subsection{Comparison of spin up vs down Permalloy}

\begin{figure}
    \centering
    \includegraphics[width=1\textwidth]{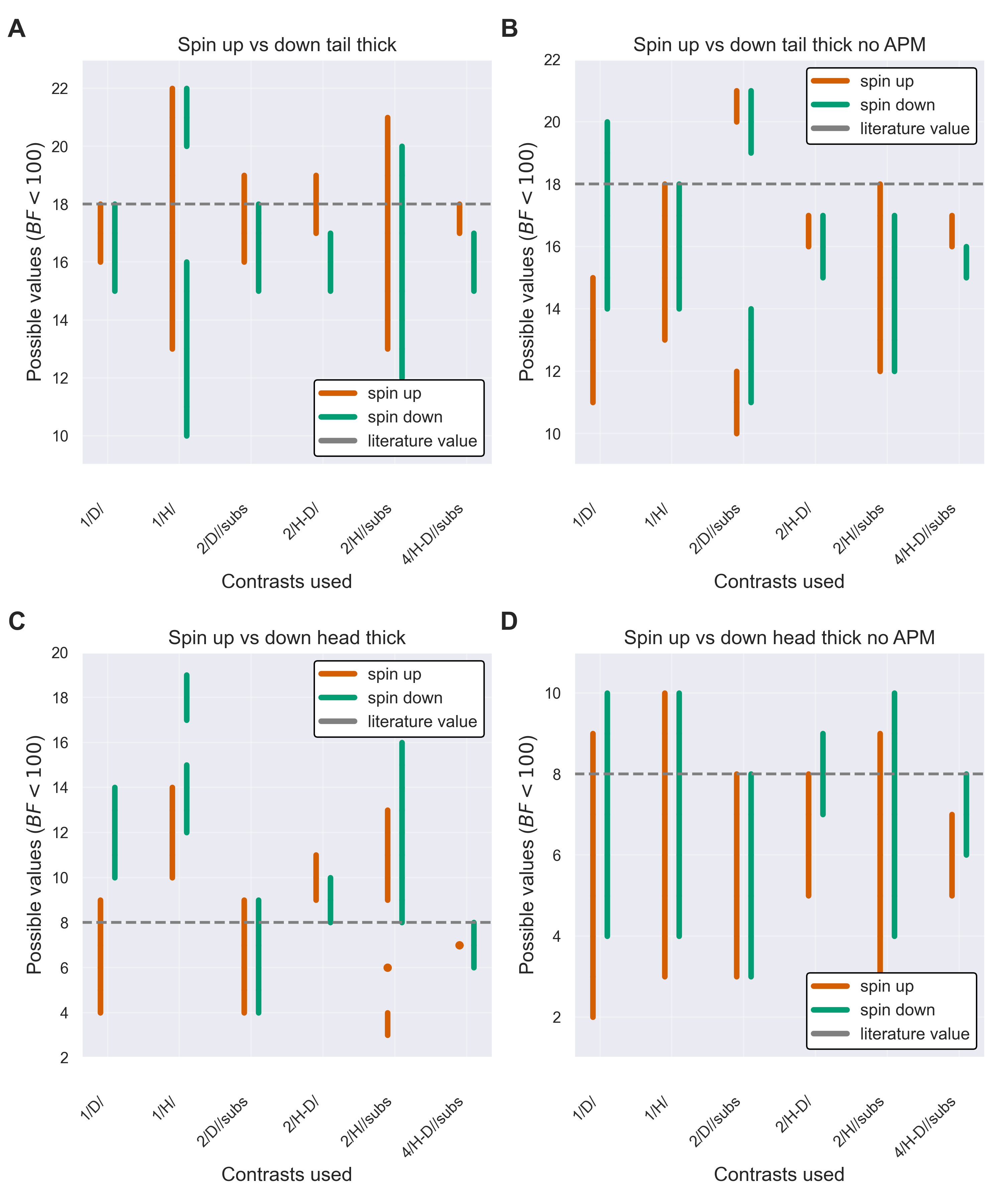}
    \vspace{-3em}
    \caption{Comparison plots of the Bayes factors below the decisive evidence value for the permalloy spin up vs down contrasts only. Indecisive evidence plot for the \textbf{(A)} bilayer tail thickness with the APM constraint and \textbf{(B)} without, \textbf{(C)} the bilayer head thickness with the APM constraint and \textbf{(D)} without.}
    \label{fig:spin-comparison}
\end{figure}

The analytical resolution for both the tail and the head thickness parameters (see Figure \ref{fig:spin-comparison} A and C) is better when the Permalloy is in the spin-up state. When the Permalloy is in the spin-up state the SLD difference between the adjacent SLD layers is greater than when the Permalloy is in the spin-down state. This implies that a greater difference in the SLDs of the substrate layers positively impacts the resolution of the parameters for the system under study. On the other hand, when the APM constraint is used (see Figure \ref{fig:spin-comparison} B and D) there is no general trend of improved resolution for either spin state.


\section{Discussion}
The findings of our study demonstrate the potential and complexities of applying Bayesian analysis in neutron reflectometry data analysis to understand the analytical resolution for a specific experimental setup and sample. Bayesian analysis enables a robust comparison of different models by quantifying the evidence supporting each one. This is achieved using Bayes factors, which provide a comprehensive measure of the relative probabilities of different models, given an equal a priori belief in the two models. By following the workflow established in this work we can get the analytical resolution for the specific data, model and prior information.

Our work confirms the significant role of contrasts in the analysis process. We found that the analytical resolution for both the tail and head thickness parameters improves as the number of contrasts increases. This intuitively makes sense, as more contrasts effectively reduce the parameter space, leading to more precise estimation. 

The Average Per Molecule (APM) assumption introduced an intriguing dynamic to the analytical resolution. While the APM assumption appears to worsen the resolution for the tail thickness parameter, it conversely improves the resolution for the head thickness parameter. This counterintuitive relationship can be attributed to the combination of uncertainties when one parameter is calculated from the other. Our findings suggest that careful consideration must be given to such model assumptions, as they can significantly impact the resolution.

We also found that the choice of contrast has a substantial effect on analytical resolution. Specifically, D2O contrasts generally provided better analytical resolution than H2O contrasts. This finding underscores the importance of careful contrast selection in neutron reflectometry studies and encourages further investigation into the properties and behaviour of these contrasts.

Finally, our study revealed a complex interplay between the permalloy spin states and the analytical resolution. Generally, the resolution was better when the Permalloy was in the spin-up state, suggesting that a greater difference in the scattering length densities of the substrate layers can positively impact the resolution. However, when the APM constraint was not used, no general trend of improved resolution was observed for either spin state. This finding invites further exploration into the interrelationship between spin states and model constraints, and how they collectively influence the resolution.

Perhaps the key takeaway of this work is that the analytical resolution in neutron reflectometry is dependent on the analysis. Model assumptions, multiple contrasts, substrate choice, and prior information all have an impact on the analytical resolution which as demonstrated above can be better than the common knowledge \ref{eqn:limit}.

In conclusion, our study illuminates the intricate factors that can impact the effectiveness and accuracy of Bayesian analysis in neutron reflectometry data analysis. By shedding light on the implications of contrast selection, model assumptions, and spin states, our findings can inform more accurate and reliable model selection in future studies. It is anticipated that continued exploration of these factors will significantly advance our understanding and application of neutron reflectometry.

\bibliographystyle{iucr} 
\bibliography{iucr} 


\end{document}